\documentclass[runningheads]{llncs}
\usepackage[T1]{fontenc}
\usepackage{graphicx}
\usepackage{url}
\usepackage{dirtree}
\usepackage{tcolorbox}
\usepackage{enumitem}
\usepackage{booktabs}
\usepackage[caption=false]{subfig}
\usepackage{bbding}
\usepackage[hidelinks]{hyperref}

\begin{document}
\title{Practical Implementation Report on Introducing Spec-Driven Development Using AI Agents in Software Development PBL}
\titlerunning{Spec-Driven Development Using AI Agents in SDPBL}
%
\author{Hidetake Tanaka\inst{1}\orcidID{0009-0003-7766-7639} \and
Hiroshi Igaki\inst{2}\orcidID{0009-0009-0810-983X}\Envelope \and
Kazumasa Shimari\inst{3}\orcidID{0000-0001-8837-5090} \and
Kiyoshi Honda\inst{2}\orcidID{0000-0002-7725-5031} \and
Naoki Fukuyasu\inst{2}\orcidID{0009-0006-8102-6692}}
\authorrunning{H. Tanaka et al.}
%
\institute{Nara Institute of Science and Technology, 8916-5 Takayama-cho, Ikoma, Nara 630-0192, Japan\\
\email{tanaka.hidetake.te0@naist.ac.jp} \and
Osaka Institute of Technology, 1-79-1 Kitayama, Hirakata, Osaka 573-0196, Japan\\
\email{\{hiroshi.igaki,kiyoshi.honda,naoki.fukuyasu\}@oit.ac.jp}\and
Wakayama University, Sakaedani 930, Wakayama, Wakayama 640-8510, Japan\\
\email{shimari@wakayama-u.ac.jp}}
\maketitle              

\begin{abstract}

In recent years, autonomous AI agents such as GitHub Copilot and Claude Code have been rapidly gaining popularity. This study reports on the practical implementation of Spec-Driven Development, a software development methodology premised on AI agents, within a Software Development Project-Based Learning (SDPBL) course for third-year undergraduate students. We defined a workflow consisting of four phases, namely investigation, planning, implementation, and review. We also established an environment tailored for the SDPBL course where AI agents generate documentation and code during each phase. We analyzed the results from three perspectives, namely students' subjective AI usage, implementation throughput, and code comprehension. The analysis reveals that AI usage patterns varied across development phases and teams. Moreover, while AI agent utilization increased implementation throughput, it also tended to encourage students to proceed with development without fully understanding the code. This study demonstrates that regular verification of code comprehension by instructors and appropriate feedback are essential for maintaining educational effectiveness when introducing SDD into SDPBL.

\keywords{Software Engineering Education \and Project-based Learning \and AI Agents \and Large Language Models \and LLMs}
\end{abstract}

\section{Introduction}
In recent years, software development utilizing Large Language Models (LLMs) and AI agents—which possess the capability to autonomously execute various tasks—has become increasingly widespread. In the field of software engineering education, the integration of LLMs into programming exercises~\cite{Guner2025,Lau2023,Liu2024} and Software Development Project-Based Learning (SDPBL) is also actively progressing~\cite{Kharrufa2026,Rasnayaka2024}.
In particular, the introduction of LLMs into SDPBL is expected to mitigate long-standing challenges, such as the gap in implementation skills among students and the disproportionate contribution of specific members to the development process~\cite{igaki2014}. On the other hand, there are still few reports on how AI agents, which provide advanced support across various stages of software development, actually impact SDPBL in practice.

The year 2025 has been called the inaugural year of AI agents, with autonomous AI tools such as Devin\footnote{\url{https://devin.ai/}}, GitHub Copilot's agent mode\footnote{\url{https://github.blog/news-insights/product-news/github-copilot-the-agent-awakens/}}, and Claude Code\footnote{\url{https://www.claude.com/product/claude-code}} being released between late 2024 and early 2025. These agentic coding assistants can autonomously perform complex tasks, including code editing across multiple files, compilation, execution, and bug fixing, toward given goals. An analysis by Li et al.~\cite{Li2025} revealed that as of June 22, 2025, AI agents were active in 61,453 repositories and had submitted 456,535 pull requests on GitHub. As AI agent adoption continues to grow, determining how to effectively and educationally incorporate them into SDPBL is expected to become a critical challenge in software engineering education.
Therefore, in this paper, we introduce Spec-Driven Development (SDD)—a development methodology designed for the use of AI agents—into an actual SDPBL course to clarify how students utilize AI agents within the development process.

Recently, SDD has been proposed as a development methodology based on AI agents\footnote{\url{https://kiro.dev/blog/from-chat-to-specs-deep-dive/}\label{fot:sdd}}. SDD involves defining detailed specifications clearly at the early stages of development and having AI agents generate code based on those specifications. This methodology requires AI agents to proceed with development based on structured specification documents. While AI agents create various specifications and documents, developers are required to understand and verify them, thereby maintaining code quality and consistency throughout the development process.

We aim to identify the realities of AI agent utilization in the context of team development education, including both benefits (such as implementation throughput improvements and development support) and potential drawbacks (such as development proceeding without adequate understanding and the burden associated with AI agent usage).

To achieve this goal, we address the following research questions:

\begin{description}[style=unboxed,leftmargin=0cm]
\item[\textbf{RQ1.}] To what extent do students utilize AI in each phase of the development process (investigation, planning, implementation, review, etc.) in SDPBL with SDD introduced?

\item[\textbf{RQ2.}] Does the widespread use of LLMs/AI agents affect students' implementation throughput (in terms of added lines of code) in SDPBL?

\item[\textbf{RQ3.}] How does the level of AI usage relate to students' code comprehension?
\end{description}

In SDPBL where SDD has been introduced, the extent to which students tend to utilize AI tools, including AI agents, in different development phases (investigation, planning, implementation, review, etc.) remains unclear. We investigate the activity status of each student and team through weekly surveys and analyze how LLMs are utilized across different phases. In these surveys, we define the level of AI usage on a five-point scale, ranging from level 0 (no use) to level 4 (very helpful), based on students' perceived effectiveness in each phase.

Next, we analyze the impact of LLM usage on implementation throughput. While it has been reported that utilizing LLMs in software development improves code contributions~\cite{Shihab2025}, other studies suggest that LLM contributions are limited to experienced developers and have limitations for inexperienced developers~\cite{stray2025}. In this study, we aim to quantitatively analyze how students, who are unfamiliar with team development in SDPBL contexts, are affected in their implementation throughput (specifically, added lines of code) by using LLMs.

Regarding the impact of AI usage, it has also been suggested that students may pay less attention to code and have limited understanding when developing with LLMs~\cite{Shihab2025}. In the SDPBL we conducted, we paid particular attention to this point and conducted one-on-one interviews with each student at each lecture session to assess their code comprehension on a four-level scale. By comparing these interview results with the previously defined levels of AI usage, we analyze how the level of AI usage relates to students' ability to understand and explain their own code.

The remainder of this paper is organized as follows. Section~\ref{sec:related_work} summarizes related work on LLM-based software development, AI agent-based software development, SDD, and generative AI in software engineering education. Section~\ref{sec:propose} details the application of SDD to our SDPBL course and its implementation status. Section~\ref{sec:results} presents our findings. Section~\ref{sec:discuss} discusses implications for introducing SDD to SDPBL and threats to validity. Section~\ref{sec:conclusion} concludes the paper.

\section{Related Work}
\label{sec:related_work}

\subsection{LLM-Based Software Development}
\label{subsec:llm}
Software development utilizing Large Language Models (LLMs) has become widespread in recent years, with a development methodology known as ``Vibe Coding'' gaining traction. This approach was popularized by AI researcher Andrej Karpathy through a post on X\footnote{\url{https://x.com/karpathy/status/1886192184808149383}\label{fot:karpathy}}, representing a concept where developers proceed with development through dialogue with generative AI without closely examining code details.
This methodology is particularly suited for rapid Proof of Concept (POC) and prototype development. Developers can quickly transform ideas into tangible forms without spending extensive time on traditional detailed implementations.

Conversely, ensuring the quality and consistency of generated code is challenging in Vibe Coding because developers do not directly verify implementation details~\cite{Tambon2025,Wang2025}.
A study by Hou et al.~\cite{Hou2024} highlights that code generated by LLMs often selects generic designs without considering project-specific coding conventions or design patterns. Furthermore, it identifies issues such as the accumulation of technical debt due to the disregard for architectural patterns.

\subsection{AI Agent-Based Software Development}
\label{subsec:agent}
AI agents have been increasingly applied to various software engineering phases.
Bouzenia et al.~\cite{RepairAgent} proposed RepairAgent, an agent that autonomously selects and executes appropriate tools for bug fixing. It iteratively performs tasks ranging from bug information collection to fixing and verification based on gathered data and feedback from previous attempts. Evaluations using the Defects4J dataset demonstrate its capability to autonomously repair 164 bugs.
Additionally, Batole et al.~\cite{LocalizeAgent} introduced LocalizeAgent to maintain design quality. This multi-agent framework summarizes program analysis results into LLM-understandable formats and identifies design issues through prompt generation tailored to specific refactoring types. It achieved up to a 206\% improvement over the baseline in localization accuracy for issues such as modularity.

According to a survey by He et al.~\cite{He2025}, AI agents are evolving beyond single-task assistance toward automating entire development processes. Their importance is expected to grow further across various software development tasks, including requirements definition~\cite{Sami2024}, design~\cite{Zhou2025}, implementation~\cite{Wang2024,PairCoder}, testing~\cite{Feldt2024}, and code review~\cite{Heander2025}.

\subsection{Spec-Driven Development}
\label{sec:sdd}
To address the challenges of Vibe Coding, Spec-Driven Development (SDD) has been proposed \footref{fot:sdd}. This approach involves defining detailed specifications clearly at the early stages of development and having AI agents generate code based on those specifications. The distinguishing feature of SDD is that AI agents proceed with development based on structured specification documents rather than ambiguous natural language instructions. These specifications explicitly describe architectural patterns, coding conventions, security requirements, and testing strategies. AI agents recognize these specifications as constraints and generate code compliant with them.

The AWS Kiro documentation\footnote{\url{https://kiro.dev/docs/specs/}}, which introduced SDD to the world in July 2025, defines a workflow where AI agents sequentially execute phases such as requirements analysis, design, and implementation planning. During each phase, documents such as \path{requirements.md}, \path{design.md}, and \path{tasks.md} are generated by AI agents. 
Developers continuously verify these documents at each step to ensure alignment with their intent. By continuously verifying and refining these specifications, developers can ensure quality consistency and maintainability throughout the subsequent implementation phase.

Many AI agents, including GitHub Copilot, can interpret these workflows and project-specific coding conventions from Markdown documents called custom instructions. For instance, by describing workflows and project-specific rules (e.g., generating outputs in Japanese or adhering to specific coding conventions) in a file named \path{copilot-instructions.md}, developers can enable AI agents to perform SDD according to the specified instructions.

In SDD, although AI agents create various specifications and documents, developers are required to understand and verify them. Consequently, SDD demands greater expertise compared to Vibe Coding. While receiving support from AI agents, developers must understand the generated documents and source code and take responsibility for the final artifacts.
Consequently, for SDPBL aimed at cultivating application development skills, SDD is more suitable than Vibe Coding. This is because students must understand the specifications and assume responsibility for the AI agent's outputs, including both documentation and source code.

\subsection{Generative AI and Software Development Education}
\label{sec:aiedu}
Surveys of university programming education instructors regarding generative AI highlight conflicting views, ranging from prohibiting tools like ChatGPT to actively integrating them with educational considerations for students' future careers~\cite{Lau2023}. Furthermore, Guner et al.~\cite{Guner2025} analyzed ChatGPT usage patterns in programming and identified five distinct student profiles (e.g., AI-dependent code generators), reporting that appropriate instructional intervention significantly improves both exam performance and usage patterns.

The integration of LLMs into team-based development, such as SDPBL, is also advancing~\cite{Kharrufa2026,Rasnayaka2024}. However, few studies have comprehensively investigated the effectiveness of AI agents when integrated into team development across various development phases. Determining how to effectively incorporate AI agents into team-based projects like SDPBL is expected to be a significant challenge in software engineering education. Therefore, this paper describes the application methodology of Spec-Driven Development (SDD) in our SDPBL course for third-year undergraduate students at Osaka Institute of Technology and presents an analysis of the students' implementation status.

\section{Application of Spec-Driven Development to SDPBL}
\label{sec:propose}
This section describes the application of SDD to our SDPBL course.
First, we provide an overview of the target course, followed by a detailed explanation of the materials, focusing on a specific case of introducing an AI agent into SDPBL within the course.
Then, we describe the evaluation methodology used throughout the course and present the approach for addressing the RQs.

\subsection{Overview of the SDPBL Course}
\label{sec:sdpbl}
This paper focuses on our SDPBL course, conducted as part of the ``Information System Development Exercise'' course for the Department of Information Systems at Osaka Institute of Technology.
This elective course is offered in the second semester of the third year and consists of 14 weekly lectures, each with two consecutive 100-minute periods (2 credits).
The objective of this PBL is for teams to develop a web application with a database, applying the knowledge of information system implementation acquired in previous coursework.
Through hands-on experiences with web frameworks, server deployment, and agile processes, students aim to acquire the essential skills and knowledge required for system engineers.
While participating students possess a foundational understanding of Java syntax and have received prior instruction on Git and GitHub, they typically have minimal experience in team-based development for Spring Boot or web applications.

The first five of the 14 lectures consist of tutorials on the Spring Boot framework.
Starting from the 6th lecture, students are divided into teams of three to four for the SDPBL, which culminates in a final presentation in the 14th lecture.
The development environment consists of Visual Studio Code, Git, and GitHub to support collaborative work.
Teams follow the GitHub Flow\footnote{\url{https://docs.github.com/en/get-started/using-github/github-flow}} and submit weekly progress reports.
These reports document each team member's development plan and implementation details for the preceding week, including branch names, corresponding pull requests (PRs), and review results.
Each lecture begins with a 30-minute retrospective using the KPT (Keep, Problem, Try) method, the results of which are included in the reports.
Following the retrospectives, instructors review the reports and interview students to provide feedback on their progress and KPT results.
Since teams are required to demonstrate a working version of their application during each weekly progress report, design and implementation activities are carried out in almost every lecture.

\subsection{Preparation and Lecture Content for Application of Spec-Driven Development}
To apply SDD to the SDPBL course, we implemented the following preparations in the Information System Development Exercise course in the 2025 academic year.

\subsubsection{Lecture on Spec-Driven Development}
\begin{figure}[tb]
	\centering
	\includegraphics[width=0.7\columnwidth]{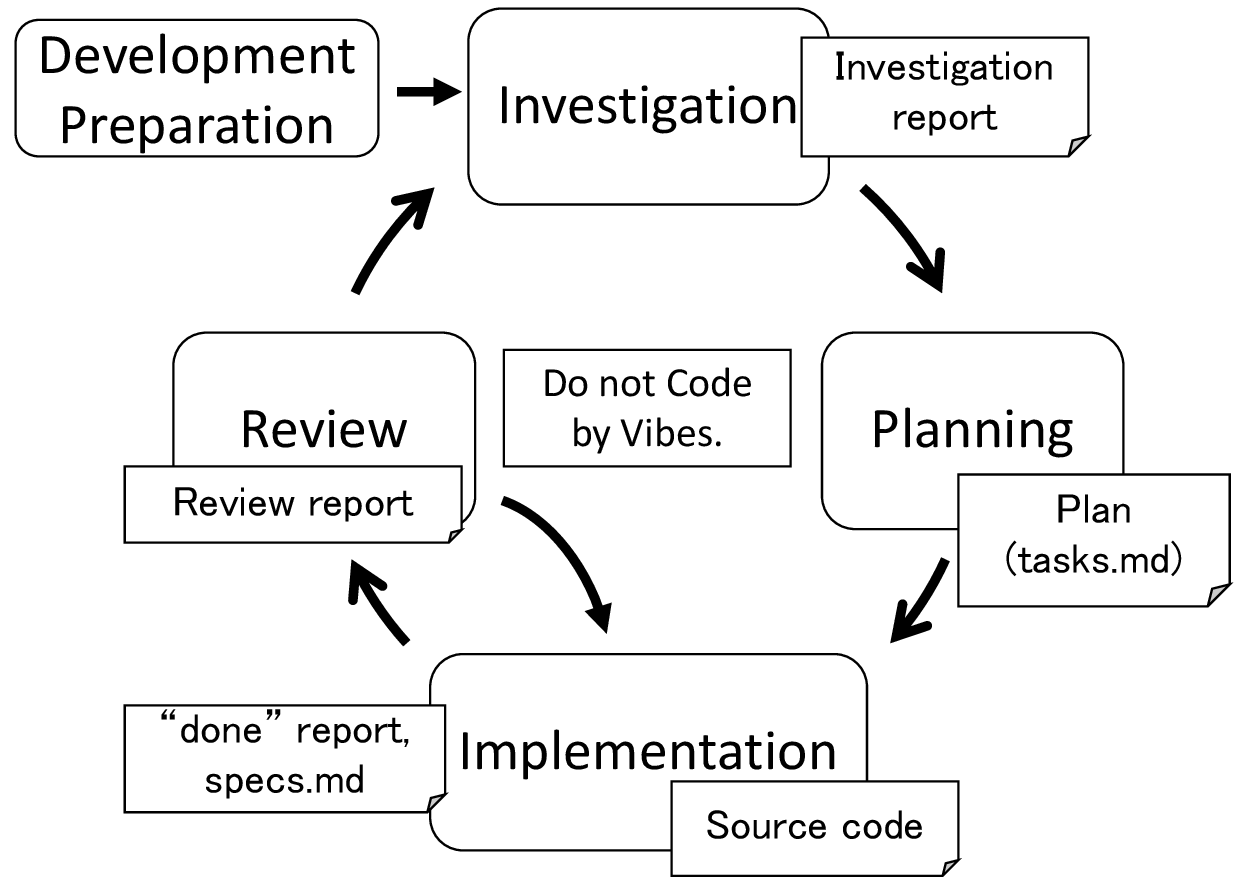}
	\caption{Workflow of Spec-Driven Development in the SDPBL course}
	\label{fig:ourworkflow}
\end{figure}

Instructors explain the concepts of GitHub Copilot (agent mode), Vibe Coding, and SDD (as discussed in Sections \ref{subsec:llm} and \ref{sec:sdd}) to the students.
Additionally, they present a workflow for applying SDD to the SDPBL course (Figure \ref{fig:ourworkflow}).
As shown in Figure \ref{fig:ourworkflow}, the development process is divided into four phases: investigation, planning, implementation, and review.
In each phase, AI agents generate artifacts, including investigation reports and implementation plans.
Specifically, the requirements analysis and design phases of the existing SDD workflow are redefined as the investigation phase in our workflow, and a review phase is added after implementation. 
This modification aims to generalize and simplify the workflow by focusing on investigating current specifications for various purposes, such as feature additions and debugging, rather than just analyzing new requirements. 
The review phase is explicitly included to have both AI agents and students confirm that the implementation matches the specifications and follows coding conventions.
Instructors emphasize the importance of carefully reviewing and understanding each AI-generated artifact to ensure consistent development.

\subsubsection{Distribution of the Template Project}
\begin{figure}[tb]
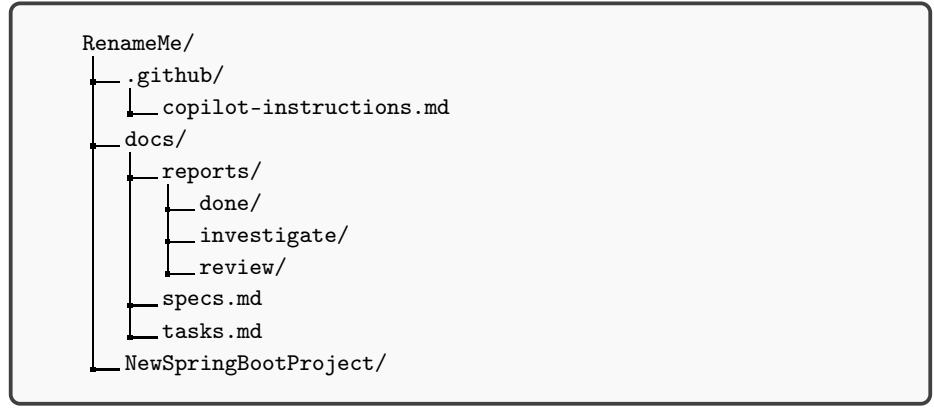

\centering
\begin{tcolorbox}[
  colback=gray!5,
  colframe=gray!50!black,
  colbacktitle=gray!20,
  coltitle=black,
  fonttitle=\bfseries\large,
  width=\columnwidth
]
\dirtree{%
.1 RenameMe/.
.2 .github/.
.3 copilot-instructions.md.
.2 docs/.
.3 reports/.
.4 done/.
.4 investigate/.
.4 review/.
.3 specs.md.
.3 tasks.md.
.2 NewSpringBootProject/.
}
\end{tcolorbox}
\caption{Directory structure of the template project}
\label{fig:dir}
\end{figure}

Following the lecture, a template project directory for Visual Studio Code is distributed to each team.
Figure \ref{fig:dir} illustrates the directory structure; AI-generated artifacts are stored in the \path{docs/} directory.
The \path{done/}, \path{investigate/}, and \path{review/} subdirectories contain the corresponding reports.
Each team renames the \path{RenameMe} directory to their project name, creates a Spring Boot project in the \path{NewSpringBootProject} directory, and registers it as a GitHub repository.
\path{copilot-instructions.md} contains the workflow and project-specific rules as custom instructions.

\subsubsection{Demo of Spec-Driven Development}
Following the explanation of the AI agents and the workflow, a demonstration was conducted where two instructors performed one full cycle of the workflow on the template project in front of the students.
The demonstration used the prompts shown in Figure \ref{fig:prompts} to guide the AI agent through the investigation, planning, and implementation phases.
The demonstration began with an investigation prompt and a review of the generated report, followed by a planning prompt to generate \path{tasks.md}.
After verifying the plan for functional requirements and testing procedures, the instructors formulated an implementation prompt.
Upon receiving the implementation instruction, the AI agent edits the source files and updates \path{specs.md}.
In the demonstration, the instructors verified that the login functionality was correctly implemented according to \path{tasks.md} and explained the verification points to the students.
The instructors emphasized that students must be able to explain any AI-generated code.
Following implementation, the AI agent reviews the code to ensure that requirements are met, coding conventions are followed, and \path{specs.md} is updated.
The resulting review report evaluates functionality, security, code quality, and documentation consistency, and provides prioritized improvement suggestions.
If issues are identified, the process returns to the investigation phase.
The project developed by the instructors with the AI agent based on this workflow is published on GitHub\footnote{\url{https://github.com/kiyoshi-honda/InuzukiProject}} (written in Japanese).

\begin{figure}[tb]
  \centering
  \begin{minipage}[t]{0.35\textwidth}
    \begin{tcolorbox}[
      colback=gray!5,
      colframe=gray!50!black,
      colbacktitle=gray!20,
      coltitle=black,
      title=Prompt for the investigation phase,
      fonttitle=\bfseries,
      width=\linewidth
    ]
    When accessing \url{http://localhost:8080/}, the login page should be displayed, and after logging in with ID yamada / PW taro, the page should be displayed. Investigate the implementation method.
    \end{tcolorbox}
  \end{minipage}
  \hfill
  \begin{minipage}[t]{0.31\textwidth}
    \begin{tcolorbox}[
      colback=gray!5,
      colframe=gray!50!black,
      colbacktitle=gray!20,
      coltitle=black,
      title=Prompt for the planning phase,
      fonttitle=\bfseries,
      width=\linewidth
    ]
    Based on the investigation results, create a plan for implementing login with in-memory user definitions (ID: yamada, PW: taro).
    \end{tcolorbox}
  \end{minipage}
  \hfill
  \begin{minipage}[t]{0.30\textwidth}
    \begin{tcolorbox}[
      colback=gray!5,
      colframe=gray!50!black,
      colbacktitle=gray!20,
      coltitle=black,
      title=Prompt for the implementation phase,
      fonttitle=\bfseries,
      width=\linewidth
    ]
    Implement the login functionality. Encode the password using BCrypt. Use the default login page for the custom login page.
    \end{tcolorbox}
  \end{minipage}
  \caption{Examples of prompts for the investigation, planning, and implementation phases}
  \label{fig:prompts}
\end{figure}

\subsection{Evaluation Methodology}
\label{sec:evaluation}
As described in Section \ref{sec:sdpbl}, the SDPBL started from the 6th lecture, and we analyze team activities from November 11, 2025 to January 13, 2026. To address RQ1--RQ3, we collect and analyze data on (1) student perception of AI (including AI agents or other LLM services) utilization (RQ1), (2) the number of lines added (LOC) by each student per lecture (RQ2), and (3) students' comprehension of the implemented content for each lecture (RQ3).

To monitor AI utilization, we conduct weekly student surveys. While our primary research focus is on AI agents, we recognize that students may employ a broad range of AI-related services beyond these specific tools. Therefore, we intentionally designed the surveys to capture the overall status of AI utilization, ensuring that any AI-assisted activities are documented regardless of the platform used. First, we assess students' AI use up to the 5th lecture—before the introduction of SDD by the instructors. From the 6th lecture onwards, the same questions are administered every lecture. Since the deadline for these surveys is set for the day before the next lecture, the responses for a given lecture (e.g., lec06) reflect activity from the day of that lecture until the day before the subsequent one.
The survey asked students to describe how they utilized Generative AI in their team development activities during the week, without restricting the scope to specific AI agents. Specifically, students rated the extent to which they used AI tools for tasks such as investigation, planning, implementation, review, debugging, and explanation (of code, etc.). Responses were provided on a 0--4 scale, where 0 indicated ``No use,'' and 1--4 ranged from ``Used but not helpful at all'' to ``Used and very helpful.'' This allowed us to analyze not only whether the tools were used but also the students' perceived effectiveness.

Regarding the number of lines added, we retrieve the Git repositories and aggregate the LOC added by each student for all commits on the \path{main} branch during each lecture's period. A lecture's period is defined from the day of the lecture until the day before the next lecture. Files that were clearly not new implementation, such as external library installations or accidentally included log files, were manually excluded by the authors.

To assess students' understanding of the code and architecture, we provided opportunities for each student to explain the code they committed to the repository during progress reports. Instructors directly assessed understanding based on a four-level scale, where 0 denotes no understanding, 1 denotes partial understanding, 2 denotes substantial understanding, and 3 denotes full understanding.

\section{Results}
\label{sec:results}

\subsection{RQ1. To what extent do students use AI in each phase of the development process in SDPBL with SDD?}
\begin{figure}[tb]
	\centering
	\includegraphics[width=\textwidth]{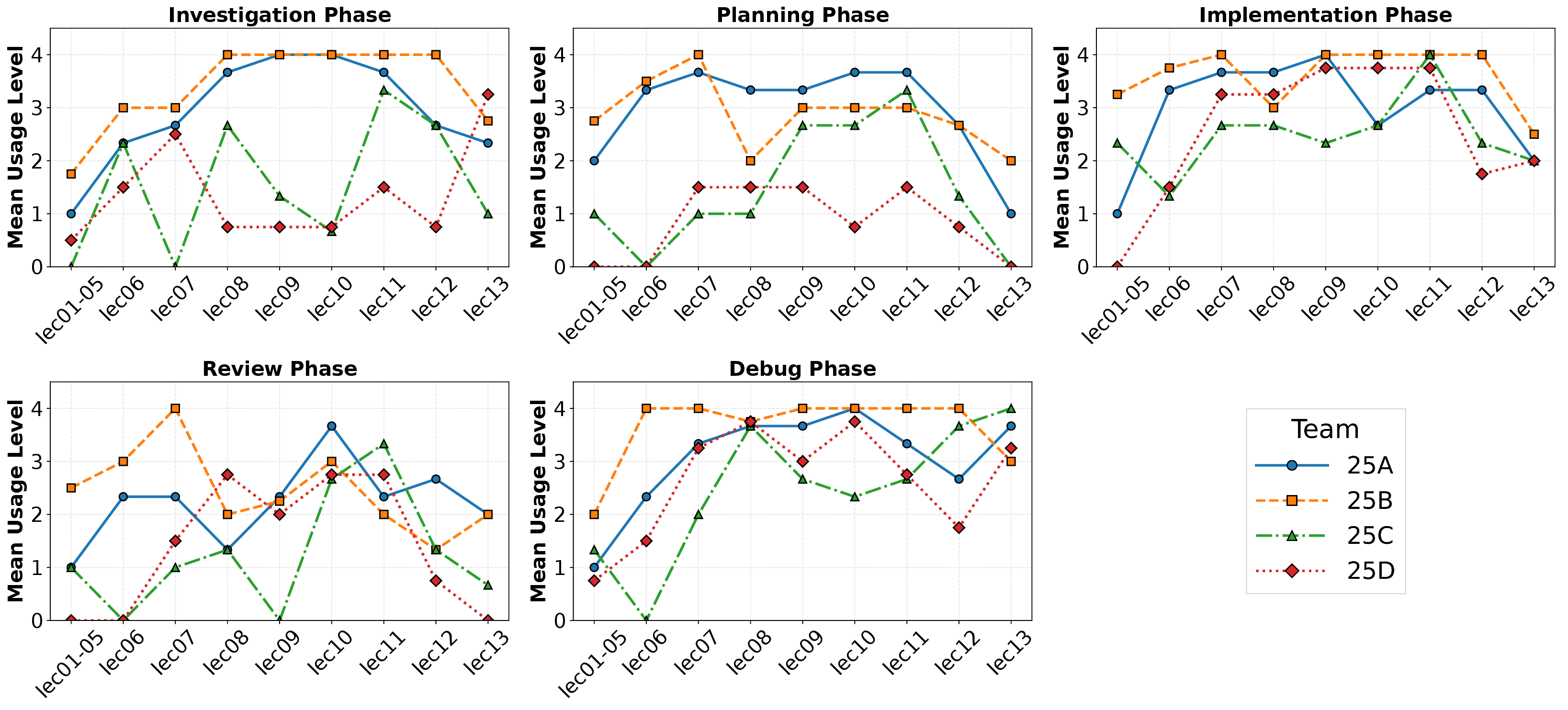}
	\caption{Trend of AI Utilization}
	\label{fig:aiuse}
\end{figure}

Each graph in Figure \ref{fig:aiuse} shows how much students used AI in each phase: Investigation, Planning, Implementation, Review, and Debugging. Students reported their AI use on a 5-point scale from 0 to 4, and these reports were aggregated by team.
lec01--lec05 show the results from the individual exercise period before team development started. These results show which phases students used AI in and how much they used it during this period. lec06 and later show the results at the end of each week. Students self-reported how much they used AI in each phase during that lecture week.

These results show, in the Implementation and Debugging phases, which relate to source code generation and understanding, all teams used AI to some extent.
However, in other phases, there are large differences in AI use between teams. In the Investigation and Planning phase, Team 25A and Team 25B used AI effectively. In the Review phase, all teams did not use AI much.
In this SDPBL, Team 25A and 25B used AI effectively throughout all phases. Team 25C used AI less compared to other teams. Team 25D showed large differences in AI use depending on the phase. Team 25D hardly used AI in Investigation and Planning. However, Team 25D used AI effectively in Implementation and Debugging. These results suggest that when SDD is introduced into SDPBL, how much students use AI in each phase is likely to differ by team. Especially in phases other than source code generation and comprehension, such as Investigation and Planning, AI use tends to differ greatly by team and by student.

\subsection{RQ2. Does the widespread use of LLMs/AI agents affect student's implementation throughput (added LOC) in SDPBL?}
ChatGPT was released in November 2022. Since that time, generative AI has been able to generate source code. However, until OpenAI announced the "Code Interpreter" function in July 2023, advanced programming support with code execution and debugging was difficult. Therefore, we estimate that generative AI was rarely used in 2022 for Spring Boot framework development, which is the target of our SDPBL.
After 2023, students who used generative AI for coding in SDPBL gradually increased. In this course, until 2024, active use of generative AI was not recommended. Also, no instruction was given to prohibit the use of generative AI.
In 2025, it became possible to use AI agent functions in GitHub Copilot. As described in Section \ref{sec:propose}, we explained SDD in the course. We recommended that students actively use AI agents for development.

Based on the above background, we conducted an analysis to evaluate the impact of LLM and AI agent use on our SDPBL. We analyzed the number of added LOC per student per lecture from 2022 to 2025. The number of students (and teams) who took our SDPBL course in each academic year was 38 (10 teams) in 2022, 28 (7 teams) in 2023, 25 (7 teams) in 2024, and 14 (4 teams) in 2025.

The data collection method is shown below:

\begin{enumerate} 
\item We collected the GitHub repository URLs of all teams that participated in SDPBL in each year. We also collected account information for each student.

\item We obtained the Git repositories of all teams. We analyzed all commits included in the period for each lecture targeting the main branch. We obtained a list of added LOC for each file included in the commits. Here, the period for each lecture is from the day of the lecture to the day before the next lecture.

\item We manually removed unnecessary files from the list of added LOC for each file. The files we removed are as follows: external library files, files mistakenly included in commits (such as log files), manuals such as README, Markdown files generated by AI agents, other files that do not affect program execution.

\item We aggregated the remaining list of added LOC for each file by student and by lecture. 
\end{enumerate}

\begin{figure}[tb]
	\centering
	\includegraphics[width=\textwidth]{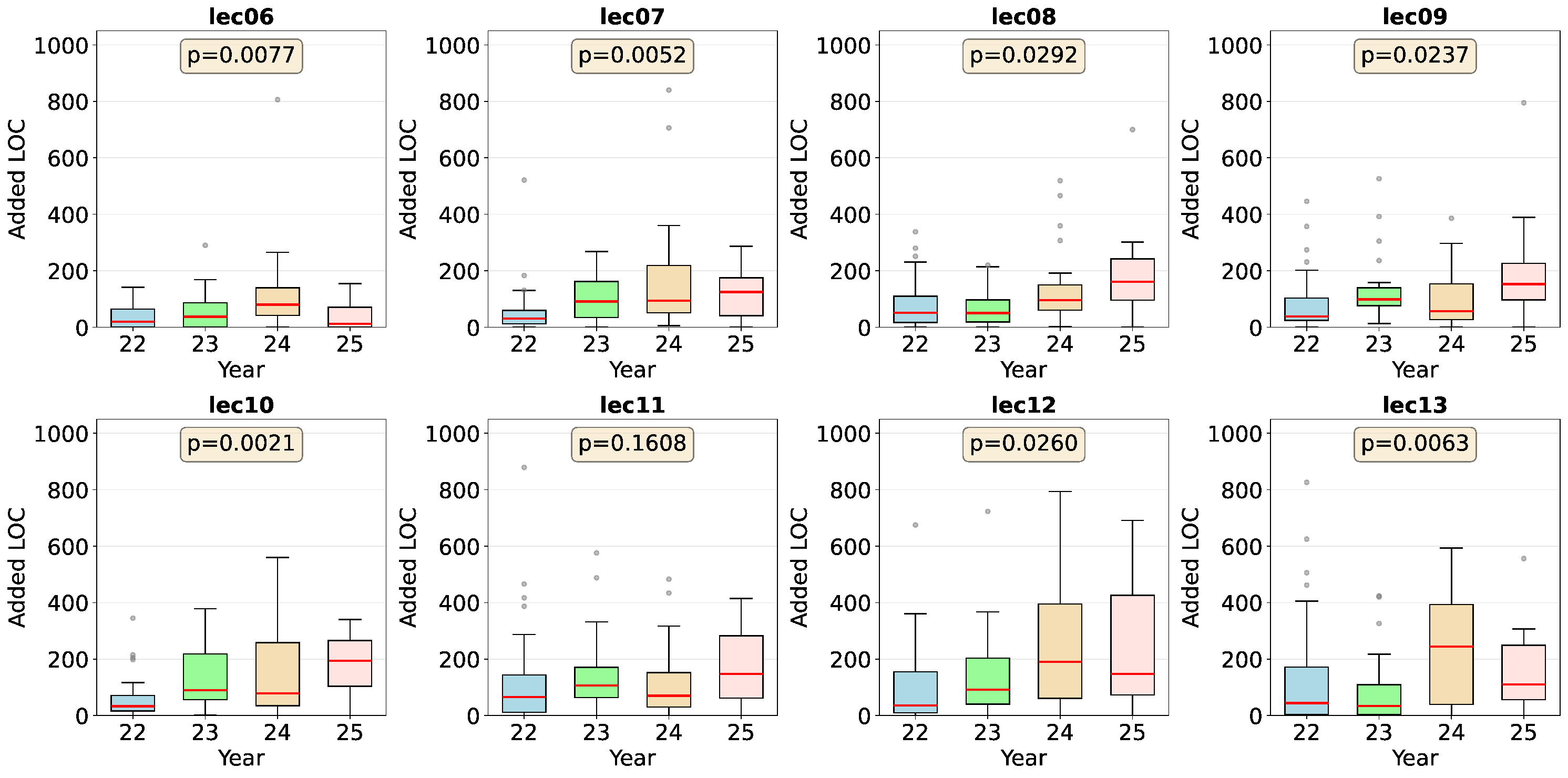}
	\caption{Year-to-year Comparison of added LOC for Each Lecture with Steel-Dwass test Results}
	\label{fig:addLines}
\small
\textit{Note:} We treated values exceeding 1000 lines as extremely large values even among outliers. We did not draw these values in the graph. The p-values shown in each graph correspond to the minimum pairwise p-value identified by the Steel-Dwass test.
\end{figure}

Figure \ref{fig:addLines} shows the number of added LOC for all students per lecture from 2022 to 2025 as box plots. This figure shows a comparison of added LOC between years. The horizontal axis of each graph shows the year. The vertical axis of each graph shows the number of added LOC.

We examined year-to-year differences in the number of added LOC during eight lectures spanning the 2022--2025 academic years. Table \ref{tab:significant_pairs} summarizes the Steel-Dwass test results; the reported \textit{p}-value is the minimum pairwise \textit{p}-value for each lecture, and the corresponding pair is highlighted in bold.
The Steel-Dwass tests revealed statistically significant differences among academic years in seven of the eight lectures: lec06, lec07, lec08, lec09, lec10, lec12, and lec13. Only lec11 showed no significant differences.
The Steel-Dwass pairwise comparisons identified specific year pairs exhibiting significant differences (Table \ref{tab:significant_pairs}). A consistent pattern is that earlier years (2022, 2023) have fewer added LOC than later years (2024, 2025). For example, Year 2022 is significantly lower than Year 2024 in lec06, lec07, lec08, lec10, and lec12, and Year 2023 is significantly lower than Year 2024 in lec08 and lec13. No instances were observed where Years 2024 or 2025 had significantly fewer added LOC than Years 2022 or 2023, suggesting an increase in students' implementation throughput (added LOC per lecture) that corresponds temporally with the proliferation of LLM technologies.

\begin{table}[tb]
\centering
\caption{Summary of Significant Year Differences by Lectures}
\label{tab:significant_pairs}
\begin{tabular}{lll}
\toprule
\textbf{lecture} & \textbf{\textit{p}-value} & \textbf{Significant Pairs} \\
\midrule
lec06 & 0.0077 & \textbf{22 $<$ 24} \\
lec07 & 0.0052 & 22 $<$ 23, \textbf{22 $<$ 24}, 22 $<$ 25 \\
lec08 & 0.0292 & 22 $<$ 24, \textbf{23 $<$ 24}, 23 $<$ 25 \\
lec09 & 0.0237 & \textbf{22 $<$ 23} \\
lec10 & 0.0021 & \textbf{22 $<$ 23}, 22 $<$ 24, 22 $<$ 25 \\
lec11 & 0.1608 & Not significant \\
lec12 & 0.0260 & \textbf{22 $<$ 24} \\
lec13 & 0.0063 & \textbf{23 $<$ 24} \\
\bottomrule
\end{tabular}

\vspace{0.3cm}
\small
\textit{Note:} \textit{p}-values are the minimum pairwise \textit{p}-values obtained from the Steel-Dwass test for each lecture. The pair shown in \textbf{bold} corresponds to this minimum \textit{p}-value. YearA $<$ YearB indicates that YearA has significantly fewer added LOC than YearB (\textit{p} $<$ 0.05, Steel-Dwass test).
\end{table}

\subsection{RQ3. How does the level of AI usage relate to students' code comprehension?}
\begin{figure}[tb]
  \centering
  
  \subfloat[Code Comprehension Trends by AI Usage Level Over Lectures\label{fig:aiuse_understand}]{%
    \begin{minipage}[t]{0.98\columnwidth}
      \centering
      \includegraphics[width=0.8\linewidth]{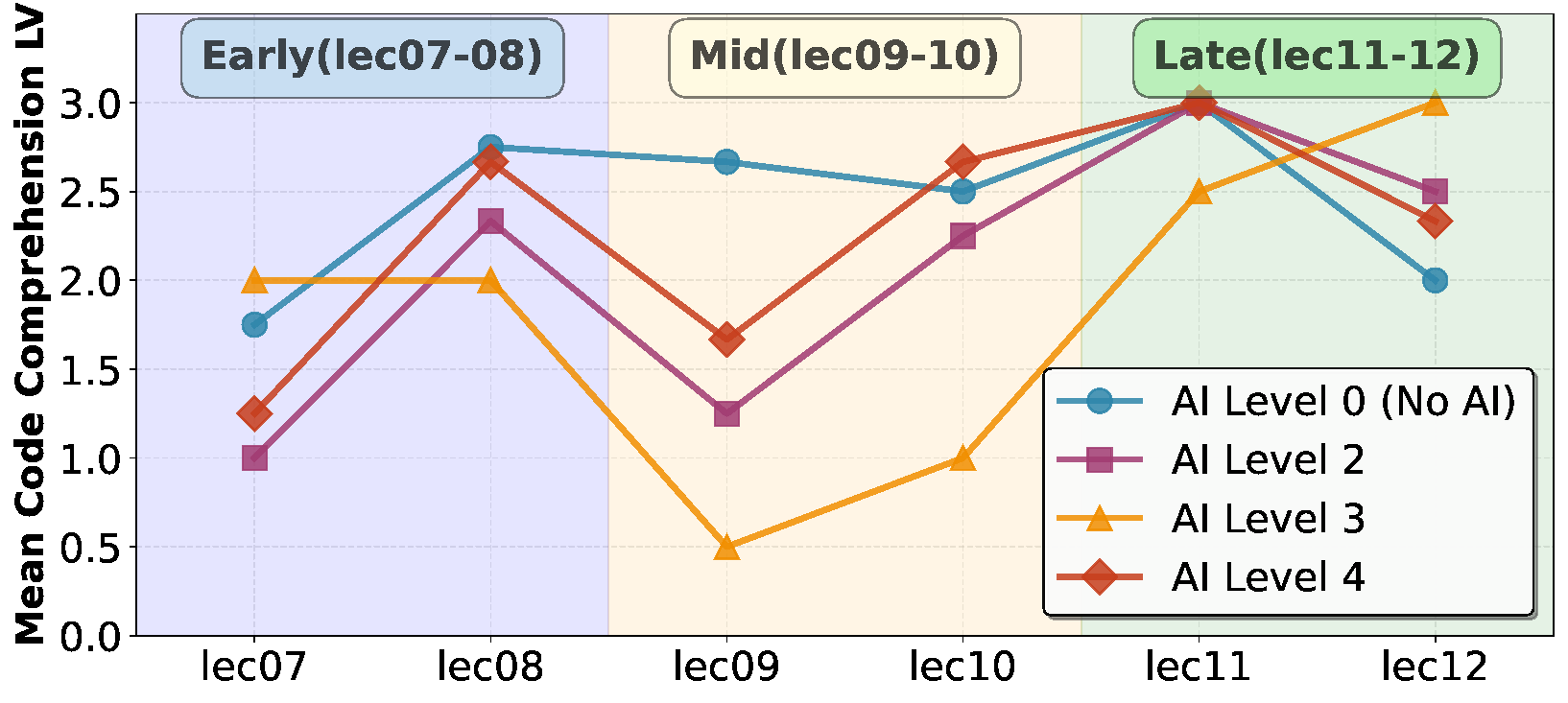}
    \end{minipage}
  }

  \subfloat[Code Comprehension Trends by Team Over Lectures\label{fig:understand}]{%
    \begin{minipage}[t]{0.98\columnwidth}
      \centering
      \includegraphics[width=0.8\linewidth]{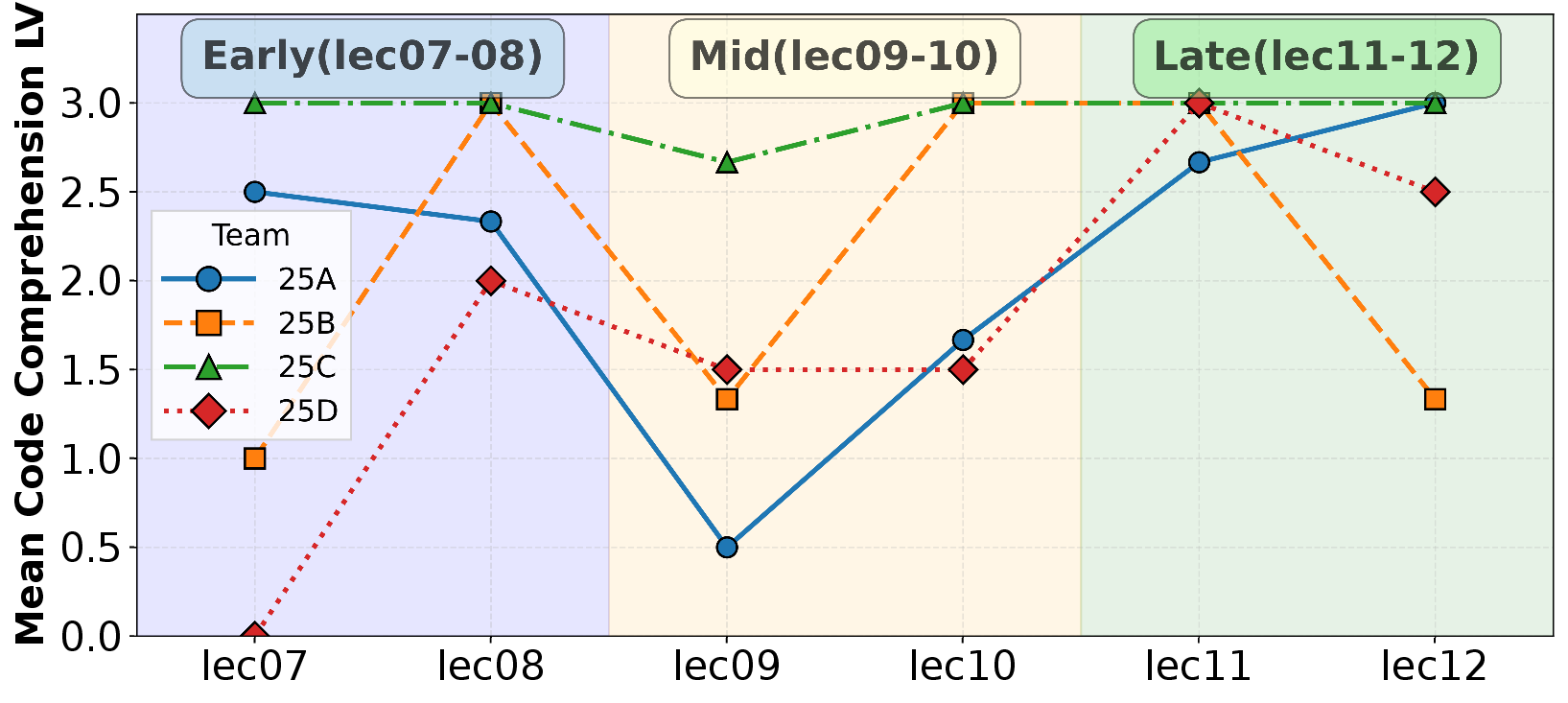}
    \end{minipage}
  }
  
  \caption{Analysis of Code Comprehension and AI Usage (Implementation Phase)}
  \label{fig:analysis}
\end{figure}

In Figure \ref{fig:aiuse_understand}, average code comprehension levels are plotted for four AI usage levels throughout the Implementation phase. Level 1 is omitted because no students reported AI usage level 1 during this phase. The analysis distinguishes three periods: Early (lec07--08), Mid (lec09--10), and Late (lec11--12). The average was calculated from each student's code comprehension level. In this figure, we can see that students at AI level 0, that is, students who implemented code without using AI, maintained consistently high code comprehension levels throughout all periods. On the other hand, students at AI level 3 implemented code with very little comprehension, especially during the Mid period (lec09--10).
Figure \ref{fig:understand} shows a graph that plots the average code comprehension level for each team across lectures. In Figure \ref{fig:understand}, we can see that Team 25C, which used AI the least during the Implementation phase, maintained consistently high code comprehension scores throughout all periods.

Furthermore, we compared the transitions of AI usage levels during the Implementation phase in Figure \ref{fig:aiuse} with the code comprehension trends in Figure \ref{fig:understand}.

By comparing these two figures, we obtained the following observations:

\begin{enumerate}
\item For all teams except Team 25C (Team 25A, Team 25B, and Team 25D), code comprehension levels fluctuated significantly across lectures. Specifically, these team’s code comprehension levels dropped in lec09. At the same time, these teams' AI usage levels during the Implementation phase (lec09) became very high.

\item We observed that most teams' code comprehension levels improved in lec08 and lec10. At this time, two patterns existed. The first pattern was teams that improved their code comprehension levels while maintaining high AI usage levels (ex. Team 25B). The second pattern was teams that improved their code comprehension levels while reducing their AI usage levels (ex. Team 25A).

\item We observed that in lec11 and lec12, most teams showed high code comprehension levels.
\end{enumerate}

In this SDPBL, instructors check each student's code comprehension level one-on-one every week from lec07 onward. Therefore, it is highly likely that students and teams pointed out by instructors for having low code comprehension tried to understand the code as much as possible in the following week.
Furthermore, we observed that code comprehension levels in the Late period (lec11--12) showed improvement trends regardless of AI usage. We believe that when introducing LLMs into SDPBL, it is important to have some form of mechanism to assess whether students understand the content of AI-generated code.

\section{Discussion}
\label{sec:discuss}
Based on our observations from the SDPBL course, this section presents practical considerations for instructors planning to introduce AI agents into SDPBL. We then discuss the threats to the validity of our findings.

\subsection{Considerations for Introducing AI Agents in SDPBL}
Based on the findings from RQ1--RQ3, we discuss three key considerations for instructors planning to adopt AI agents in SDPBL.

\subsubsection{Balancing Code Comprehension and AI Usage}
RQ1 and RQ3 explored the relationship between code comprehension and the utilization level of AI. Our observations indicate that students who did not use AI at all tended to exhibit high code comprehension scores. However, across the student body as a whole, no clear statistical trend was confirmed between AI usage levels and code comprehension.

Several factors likely contributed to this result. First, the limited sample size may have reduced the statistical power to detect a definitive trend. Second, the Standard Error of the Mean (SEM) was relatively large across all levels, reflecting high individual variability in how students integrated AI into their learning. Furthermore, the results may have been influenced by timely instructor interventions, such as one-on-one interviews intended to provide targeted guidance to students with lower comprehension levels. A potential example of this impact was observed in lec09, where AI usage peaked and comprehension scores dropped; however, following instructor support, comprehension showed a measurable recovery in lec10. These interventions may have encouraged students to prioritize logic over simple code generation, potentially mitigating the risk of over-reliance on AI and narrowing the usage gap across proficiency levels.

Meanwhile, the RQ2 results indicate an increase in implementation throughput (added LOC per student) from 2022 to 2024--2025. This trend suggests that AI may accelerate the pace of development. However, this increased output does not necessarily imply deeper learning, as students might prioritize task completion over a thorough understanding of the produced code. Balancing this increased implementation speed with the necessity of ensuring fundamental code comprehension remains a key pedagogical consideration.

We recommend three approaches to address these challenges:

\begin{itemize}

\item \textbf{Foundation-first approach:} In our SDPBL, the first five of the 14 sessions were designated as a tutorial period, covering the fundamentals of Web application development and team collaboration. By providing reference sample applications, we aimed to support students in understanding and implementing code independently. This foundation appears to be essential for building basic skills before students begin to utilize AI tools more extensively.

\item \textbf{Continuous monitoring with timely intervention:} Regular assessments are necessary to detect comprehension drops early. As demonstrated in lec09, proactive intervention is effective when applied promptly after identifying that AI usage is replacing rather than supporting comprehension.

\item \textbf{Accountability for AI-generated code:} Although a significant negative correlation between AI usage and code comprehension was not observed, the instructor's one-on-one interviews may have played a role in maintaining student engagement. These interviews can serve as an opportunity for students to demonstrate accountability for their work. Rather than using AI as a shortcut, students are encouraged to explain the logic behind any code they did not manually write, which may help ensure that implementation speed does not compromise their underlying understanding.

\end{itemize}

\subsubsection{Technical Infrastructure}
To support the integration of AI agents into the SDPBL, we distributed standardized project templates followed by a 30-minute lecture and a one-hour demonstration. Based on our observations, these project templates appeared to help students smoothly initiate their development tasks using AI agents.

These standardized templates should include pre-configured AI instructions (\path{copilot-instructions.md}), specification templates, and development rules, such as coding conventions. Such tools may reduce cognitive load, potentially allowing students to focus on high-level design and learning rather than environment setup. Moreover, the provided templates appear to serve dual purposes: for students with less experience, they offer concrete examples of how to structure instructions for AI agents; for more motivated students familiar with AI tools, they may provide a foundation for experimenting with customization and refinement. In this way, these materials can serve as valuable learning resources adapted to different skill levels.

\subsection{Threats to Validity}
\label{sec:threats}
\textbf{Internal Validity:} 
In this study, we did not conduct a controlled experiment with a concurrent control group that abstained from using AI agents. Although we compared the results with historical data from 2022 to 2024, we cannot rule out the possibility that year-to-year variations in student skill sets, differences in cohort size, or subtle changes in the educational environment served as confounding factors. Therefore, care must be taken in concluding that the observed changes are solely attributable to the introduction of AI agents.

\textbf{Construct Validity:} 
We used ``added LOC'' as the primary metric for development efficiency. However, efficiency in software engineering is a multifaceted concept that encompasses class design quality, refactoring, and the removal of redundant code. Since added LOC cannot fully capture these qualitative activities, the results may not entirely reflect the comprehensive nature of software development efficiency. 
Additionally, the code comprehension scores rely on instructors' subjective evaluations. While this approach lacks the rigor of objective tests, it enables the assessment of the depth of students' understanding through dialogue, which is difficult to capture with purely quantitative metrics.

\textbf{External Validity:} 
This study focuses on a specific course using Java and Spring Boot and is limited to an experiment involving only four teams. Consequently, the results cannot be fully generalized to all contexts. However, they provide relevant and informative guidance for similar PBL courses and team developments that employ widely adopted technologies.

\section{Conclusion}
\label{sec:conclusion}
This study examined the integration of AI agents into our SDPBL course, analyzing student activities through self-reported AI usage, added lines of code, and code comprehension levels.
Our findings provide several key insights. Students actively utilized AI agents across all development phases, with usage patterns varying between teams. We observed a notable increase in implementation throughput, with students in 2024--2025 adding more lines of code compared to those in 2022--2023. Regarding the relationship between AI usage and code comprehension, although mean values suggest that students with lower comprehension might rely more on AI, this trend was not statistically significant. The lack of a significant correlation may be attributed to our pedagogical approach, where one-on-one instructor interviews and timely interventions potentially encouraged students to maintain their comprehension of AI-generated code, thereby mitigating the risk of over-reliance.

We conclude that while AI agents can enhance implementation speed in SDPBL, their introduction should be accompanied by careful educational management. Regular monitoring of code comprehension and proactive feedback appear to be important for maintaining educational effectiveness. Providing proper guidance may help ensure that students use AI as a supportive tool rather than a substitute for deep understanding.

Future work should focus on three areas. First, developing effective methods to assess and improve students' understanding of AI-generated code. Second, investigating optimal workflows that balance AI assistance with learning outcomes. Third, conducting longitudinal studies to examine how early exposure to AI agents affects students' long-term software engineering skills.

\begin{credits}
\subsubsection{\ackname} This work has been supported by JST BOOST, Japan Grant Number JPMJBS2423, JSPS KAKENHI Grant Number 26K15097 and 26K21197, and Support Center for Advanced Telecommunications Technology Research.

\subsubsection{\discintname} The authors have no competing interests to declare that are relevant to the content of this article.
\end{credits}
%
%
%
\bibliographystyle{splncs04}
\bibliography{mybibliography}
\end{document}